\documentclass[twocolumn,showpacs,preprintnumbers,amsmath,amssymb]{revtex4}
\usepackage[english]{babel}
\usepackage[latin2]{inputenc}
\usepackage{epsfig}

\begin{document}

\title{Compositional Dependence of Formation Energies of Substitutional
and Interstitial Mn in Partially Compensated (Ga,Mn)As}

\author{J.~Ma\v{s}ek$^{a}$, I.~Turek$^{b,c}$,
J.~Kudrnovsk\'{y}$^{a}$, F.~M\'{a}ca$^{a}$, and V.~Drchal$^{a}$ }

\address{$^{a}$Institute of Physics, AS CR, Na Slovance 2, 182 21
Prague 8
\\ $^{b}$Institute of Physics of Materials, AS CR, \v{Z}i\v{z}kova
22, 616 62 Brno \\ $^{c}$Faculty of Mathematics and Physics,
Charles University, Ke Karlovu 5, 121 16 Prague}


\begin{abstract}
We use the density-functional theory to calculate the total energy
of mixed crystals (Ga,Mn)As with a small concentration of various
donors. We find that the formation energy of Mn depends strongly
on the partial concentrations of Mn in the substitutional and
interstitial positions, and on the concentration of other dopants.
The composition dependence of the formation energies represents an
effective feedback mechanism, resulting in the self-compensation
property of (Ga,Mn)As. We show that the partial concentrations of
both substitutional and interstitial Mn increase proportionally to
the total concentration of Mn.
\end{abstract}

\maketitle

\section{Introduction}

The dilute magnetic semiconductors (DMS), such as GaAs doped with
a large amount of Mn, represent an important class of mixed
crystals with promising applications in spin electronics
\cite{Ohno03}. The ferromagnetic behavior of these materials is
mediated by the holes in the valence band \cite{Dietl97,Dietl00}.
It is sensitive to the number of free carriers and to the level of
charge compensation. Mn atoms substituted in the cation sublattice
of a III-V semiconductor are acceptors and produce one hole each.
It is known, however, that some Mn atoms occupy the interstitial
positions and act as double donors
\cite{Yu02,Masek01,Maca02,Sanvito02,Blinowski03}.

The interplay between the substitutional and interstitial
incorporation of Mn into the GaAs lattice has, together with
co--doping (cf. Ref. \cite{Yu03}), a crucial effect on the
physical properties of the mixed crystal (Ga,Mn)As. A systematic
study of formation energies of the substitutional (Mn$_{\rm Ga}$)
and interstitial (Mn$_{\rm int}$) manganese can help to understand
it on a microscopic level.

In the case of weak doping small changes in the impurity
concentration can easily move the Fermi energy $E_{F}$ across the
band gap with a negligible influence on the density of states.
That is why the dependence of the formation energies on the number
of electrically active impurities is usually represented by their
dependence on $E_{F}$. The Fermi-level dependent formation energy
is obtained by adding (or subtracting) $\Delta E_{F}$ to the
formation energy calculated for a particular electronic
configuration \cite{Zunger03}. In the case of a strongly doped and
mixed crystals, however, the redistribution of the electron states
in the valence band due to the impurities cannot be neglected and
the density-of-states effect modifies the simple Fermi-level rule
for the formation energies.

That is why the knowledge of the formation energy as a function of
the impurity concentrations is necessary in the DMS. To calculate
the energy needed to incorporate Mn and other impurities in a
mixed crystal, we use the trick relating them to the
composition-dependent total energy of the mixed crystal
\cite{Masek02}. This quantity is obtained within the
density--functional theory for a series of (Ga,Mn)As mixed
crystals with various content of Mn in substitutional and
interstitial positions, and with variable concentration of the
compensating donors. The use of the coherent potential
approximation (CPA) combined with the tight--binding
linearized--muffin--tin--orbital method (TB-LMTO) \cite{Turek97}
makes possible to change the chemical composition continuously.
The lattice relaxation around the impurities and the clustering of
the Mn atoms are omitted within the CPA. For simplicity, we
consider only the interstitial Mn atoms in the T(As$_{4}$)
position; the energy of the other T(Ga$_{4}$) position is almost
the same \cite{Masek04}. The formation energies are obtained as
the first derivatives of the total energy with respect to the
corresponding partial concentration \cite{Masek02}.

Assuming a quasi--equilibrium deposition conditions, characterized
by an effective growth temperature, we use the calculated
formation energies to estimate the numbers of Mn$_{\rm Ga}$ and
Mn$_{\rm int}$ in (Ga,Mn)As mixed crystal. We also present a
simple way to determine the partial concentrations directly from
the composition dependence of the formation energies, without
solving thermodynamical balance equations.

\section{Composition dependence of the formation energies}

\begin{figure}
\begin{center}
\includegraphics[width=50mm,height=90mm,angle=270]{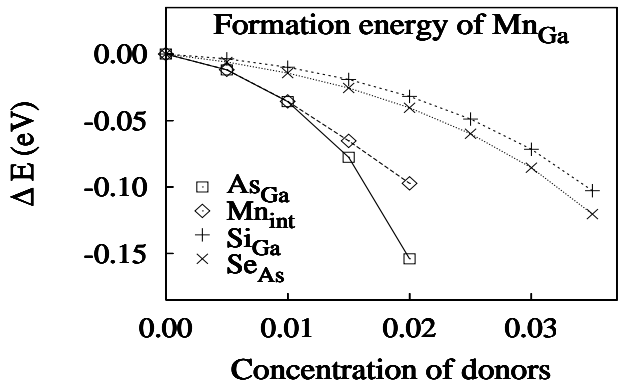}

\caption{Relative formation energy $\Delta E$ of the
substitutional Mn$_{\rm Ga}$ in Ga$_{0.96}$Mn$_{0.04}$As as a
function of the concentration of various donors.}
\end{center}
\end{figure}

We consider an impure or mixed crystal with several kinds of
impurities I$_{1}$, I$_{2}$, etc.. The total energy of the mixed
crystal $W(x_{1},x_{2}, ..)$, normalized to a unit cell, depends
on their molar concentration $x_{i}$. As we showed recently
\cite{Masek02}, the formation energy $E_{i}$ of an impurity
I$_{i}$ can be obtained by differentiating $W(x_{1},x_{2}, ..)$
with respect to $x_{i}$, namely
\begin{equation}
E_{i}(x_{1},x_{2}, ..) = \frac{\partial W(x_{1},x_{2},
..)}{\partial x_{i}} - E^{atom}({\rm I}_{i}) + E^{atom}({\rm
host}) ,
\end{equation}
The last two terms in Eq.~(1) are the total energies of an
atom I$_{i}$ and of the corresponding atom of the host, which has
been replaced by I$_{i}$.

Generally, the definition of the formation energy is not unique
and depends on the way in which the atomic energies $E^{atom}({\rm
host})$ and $E^{atom}({\rm I}_{i})$ in Eq. (1) are obtained. We
use the energies of neutral atoms in their ground state. It is
important to notice, however, that the additional constant in Eq.
(1) does not depend on the actual chemical composition of the
material. It is not important for the concentration--dependent
trends we have in mind.

That is why we consider now only the relative formation energies
$\Delta E_{i}$, obtained from their actual values $E_{i}$ by
subtracting the corresponding formation energy calculated for the
reference material. As a reference, we take
Ga$_{0.96}$Mn$_{0.04}$As with all Mn atoms in regular Mn$_{\rm
Ga}$ positions.

The composition dependence of the formation energies $E_{i}$ is
characterized by the coefficients of the linear expansion around
the reference point,
\begin{equation}
K_{ij} \equiv \frac{\partial E_{i}}{\partial x_{j}} =
\frac{\partial ^{2}W(x_{1},x_{2}, ..)}{\partial x_{i}
\partial x_{j}}.
\end{equation}
The correlation energies $K_{ij}$ \cite{Masek02} form a symmetric
matrix. $K_{ij}<0$ means that the presence of the defects I$_{i}$
supports formation of I$_{j}$ and {\sl vice versa}; $K_{ij}>0$
indicates the opposite tendency.

\begin{figure}
\begin{center}
\includegraphics[width=50mm,height=90mm,angle=270]{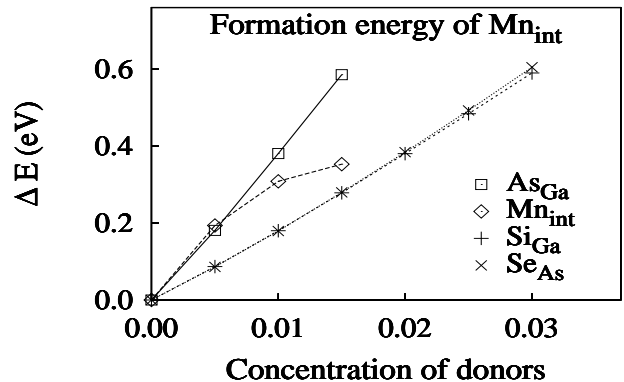}

\caption{Relative formation energy $\Delta E$ of the interstitial
Mn in Ga$_{0.96}$Mn$_{0.04}$As as a function of the concentration
of various donors.}
\end{center}
\end{figure}

The dependence of the relative formation energy $\Delta E$ of the
Mn$_{\rm Ga}$ on the concentration of various donors is summarized
in Fig. 1. We considered four representative examples. Se$_{\rm
As}$ and Si$_{\rm Ga}$ are typical donors with one extra electron,
situated at anion and cation sublattice, respectively. The other
two cases, i.e., As antisite defect As$_{\rm Ga}$ and Mn$_{\rm
int}$ are the most important native defects in (Ga,Mn)As, both
acting as double donors. Fig. 1 shows that the formation energy of
Mn$_{\rm Ga}$ decreases in the presence of an increasing number of
donors. The curves are grouped into pairs according to the charge
state of the donors, with only a minor influence of the particular
chemical origin of the defect. The dependence is almost linear for
low concentrations and the slope of the function is roughly
proportional to the charge state of the donor. This all indicates
that the variations of the formation energy of Mn$_{\rm Ga}$ are
mostly determined by the Fermi--level effect, not by the
redistribution of the density of states induced by the other
defects.

Analogous results are obtained for the formation energy of the
interstitial Mn in the T(As$_{4}$) position, as shown in Fig. 2.
In this case, however, the formation energy of Mn$_{\rm int}$
increases with increasing number of the donors. It is important to
notice that the steep increase of the formation energy represents
a feedback mechanism limiting efficiently the number of Mn$_{\rm
int}$. The same is valid also for the formation energy of the As
antisite defect.

Fig. 3 shows the formation energies of the two native defects,
i.e. As$_{\rm Ga}$ and Mn$_{\rm int}$ , in the mixed crystal with
varying number of the substitutional Mn in the Ga sublattice.
Notice that both relative quantities, being pinned to zero for the
reference material with 4 \% of Mn$_{\rm Ga}$, are almost
identical. In both cases, the formation energy is a decreasing
function, indicating an increasing probability of formation of
these defects in materials with a higher concentration of Mn.

This self-compensation tendency is a very important mechanism
controlling the basic physical properties of (Ga,Mn)As mixed
crystals.  It is the reason for the observed low doping efficiency
of Mn in GaAs \cite{Beschoten99}. The increasing number of both
As$_{\rm Ga}$ and Mn$_{\rm int}$ also explains the expansion of
the lattice of (Ga,Mn)As with an increasing concentration of Mn
\cite{Masek03}.

\begin{figure}
\begin{center}
\includegraphics[width=50mm,height=90mm,angle=270]{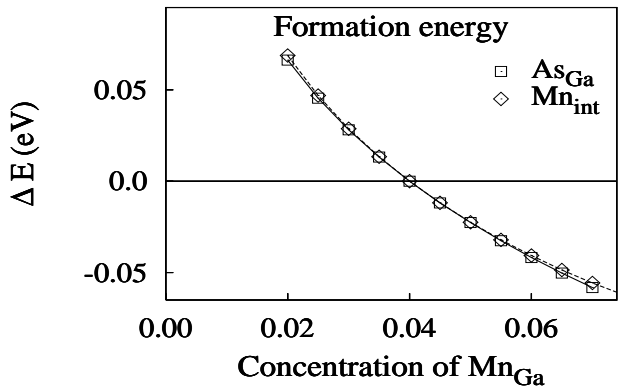}

\caption{The formation energy of the interstitial Mn and As
antisite defect in Ga$_{1-x}$Mn$_{x}$As as a function of the
concentration $x$ of Mn atoms substituted in the Ga sublattice.
The formation energies are referred to the values corresponding to
Ga$_{0.96}$Mn$_{0.04}$As}
\end{center}
\end{figure}

\section{Dynamical equilibrium between M\lowercase{n}$_{\rm Ga}$
and M\lowercase{n}$_{\rm int}$}

In this Section, we use the calculated formation energies to
simulate the incorporation of Mn into the (Ga,Mn)As mixed crystal.
We assume that the probabilities that an Mn atom occupies either
substitutional or interstitial position are related to the
corresponding formation energies $E_{S}$ and $E_{I}$ also in the
non-equilibrium epitaxial growth. As a simplest approximation, we
characterize the deposition condition with some effective
temperature $T_{eff}$ and use the corresponding Boltzmann
weighting factors.

To this purpose, the absolute formation energies of Mn in the two
crystallographic positions, i.e. Mn$_{\rm Ga}$ and Mn$_{\rm int}$,
are required. They are obtained for the reference system
Ga$_{0.96}$Mn$_{0.04}$As including the additive terms from Eq.
(1). For the epitaxial growth, it is reasonable to use $E^{atom}$
calculated for isolated atoms as stated above. A linear
interpolation for the dependence of $E_{S}$ and $E_{I}$ on the
corresponding partial concentrations $x_{S}$ and $x_{I}$ is used:
\begin{equation}
E_{S}(x_{S},x_{I}) = E_{S}^{o} + \kappa_{SS}x_{S} +
\kappa_{SI}x_{I}
\end{equation}
\begin{equation}
E_{I}(x_{S},x_{I}) = E_{I}^{o} + \kappa_{IS}x_{S} +
\kappa_{II}x_{I}
\end{equation}
with $E_{S}^{o}=0.31$~eV, $E_{I}^{o}=0.42$~eV,
$\kappa_{SS}=-0.17$~eV, $\kappa_{SI}=\kappa_{SI}=-6.03$~eV, and
$\kappa_{II}=10.33$~eV.

\begin{figure}
\begin{center}
\includegraphics[width=55mm,height=90mm,angle=270]{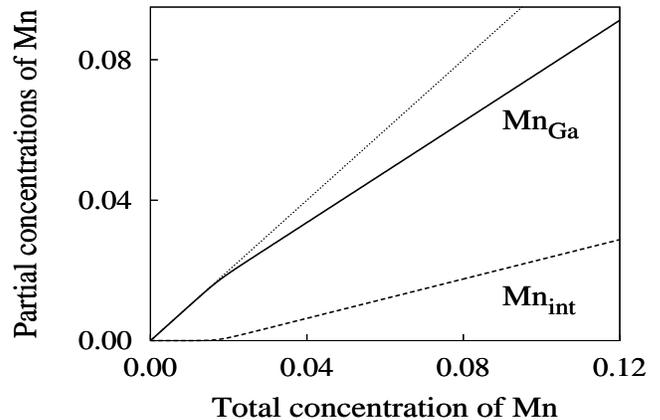}

\caption{The partial concentrations of Mn$_{\rm Ga}$ (solid) and
Mn$_{\rm int}$ (dashed) as a function of the total concentration
of Mn in (Ga,Mn)As for $T_{eff}=500$~K. The influence of other
dopants is not considered.}
\end{center}
\end{figure}

The composition dependent formation energies, Eqs. (3-4), define
the thermodynamic probabilities $p_{S}$ and $p_{I}$ that extra Mn
atoms occupy substitutional or interstitial position in the mixed
crystal with a given composition. They are
\begin{equation}
p_{S, I} = \frac{\exp(-E_{S, I}/kT_{eff})}{\exp(-E_{S}/kT_{eff}) +
\exp(-E_{I}/kT_{eff})}.
\end{equation}
On the other hand, these probabilities determine the number of Mn
atoms that substitute for Ga or occupy the interstitial positions.
The resulting changes of the partial concentrations $x_{S}$ and
$x_{I}$ due to the variation $dx$ of the total concentration $x$
of Mn are
\begin{equation}
dx_{S} = p_{S} dx, ~ dx_{I} = p_{I} dx,
\end{equation}
and the dependence of $x_{S}$ and $x_{I}$ on $x$ can be obtained
by the integration of Eq. (6).

Fig. 4 shows the solution of Eq.(6) for $T_{eff} = 500$~K. For the
lowest concentrations ($x < 0.015$), Mn atoms occupy
preferentially the substitutional positions which have lower
formation energy (cf. Eqs.(3,4)). For higher concentration of Mn,
however, the difference of $E_{I}(x_{S},x_{I})-
E_{S}(x_{S},x_{I})$ decreases and approaches zero. From this
point, both positions can be occupied with a comparable
probability, and the partial concentrations of both Mn$_{\rm Ga}$
and Mn$_{\rm int}$ increase proportionally to $x$. Any deviation
from the situation with $E_{S} = E_{I}$ changes the formation
energies of Mn$_{\rm Ga}$ and Mn$_{\rm int}$ in such a way that
the dynamical equilibrium is restored. As a result, the
high-concentration regime with co-existing Mn$_{\rm Ga}$ and
Mn$_{\rm int}$ is stabilized. This finding does not depend much on
$T_{eff}$ over a wide temperature range. We can conclude that the
partial concentrations of Mn are simply given by the equation
\begin{equation}
E_{S}(x_{S},x_{I}) = E_{I}(x_{S},x_{I}),
\end{equation}
together with the condition $x_{S} + x_{I} = x$. Combining Eqs.
(3,4) with (7) we find that the proportion of Mn$_{\rm Ga}$ and
Mn$_{\rm int}$ atoms is roughly 3:1, in a very good agreement with
both experiment \cite{Yu02} and theoretical expectations
\cite{Maca02, Masek03} for the as--grown materials.

It is important to point out that according to Eqs. (3,4,7) both
Mn$_{\rm Ga}$ and Mn$_{\rm int}$ atoms remain metastable with the
activation energy $\approx 0.3$~eV in the whole concentration
range shown in Fig. 4. The formation energies used in our
dynamical--equilibrium approach control the preferential
incorporation of Mn atoms during the growth. The annealing
process, on the other hand, depends on the barriers preventing the
Mn atoms to leave their metastable positions. The barriers are
lower for Mn$_{\rm int}$ than for Mn$_{\rm Ga}$ position
\cite{Erwin02}, so that the post--growth treatment can
substantially reduce the number of the interstitial Mn atoms
without a remarkable change of the number of Mn$_{\rm Ga}$.

\section{Summary}

We have shown that the formation energies of Mn in either
substitutional or interstitial position depend strongly on the
partial concentrations of both Mn$_{\rm Ga}$ and Mn$_{\rm int}$,
and also on the number of compensating donors. Also the formation
energy of As$_{\rm Ga}$ antisite, the main native defect in
(Ga,Mn)As, is very sensitive to the concentration of Mn.

The composition dependence of the formation energies represent a
feedback mechanism which defines a dynamical equilibrium between
Mn$_{\rm Ga}$, Mn$_{\rm int}$, and other defects and impurities
during the growth. In particular, we found that at higher Mn
concentrations the number of both Mn$_{\rm Ga}$ and Mn$_{\rm int}$
increases proportionally to the total concentration of Mn in the
as--grown (Ga,Mn)As mixed crystal.

In addition, the concentration dependence of the formation energy
of the As$_{\rm Ga}$ antisite defects indicates that an increasing
number of these donors also participate in the compensation of the
regular Mn$_{\rm Ga}$ acceptors for higher Mn concentrations.\\

\noindent {\bf Acknowledgment} This work has been done within the
project AVOZ1-010-914 of the AS CR. The financial support was
provided by the Academy of Sciences of the Czech Republic (Grant
No. A1010214) and by the Grant Agency of the Czech Republic
(202/04/583).

\end{document}